\documentclass{elsart}
\usepackage{amsfonts}
\usepackage{amsmath}
\usepackage{amssymb}
\usepackage{epsfig}
\usepackage{subfigure}
\usepackage{longtable}
\usepackage{epsf}
\usepackage{color}
\begin{document}
\runauthor{PKU}
\begin{frontmatter}
\title{Hadron Scattering in an Asymmetric Box\thanksref{fund}}

China Lattice QCD Collaboration (CLQCD)\\

 \author[PKU]{Xin Li},
 \author[IHEP]{Ying Chen},
 \author[PKU]{Guo-Zhan Meng},
 \author[PKU]{Xu Feng},
 \author[PKU]{Ming Gong},
 \author[PKU]{Song He},
 \author[IHEP]{Gang Li},
 \author[PKU]{Chuan Liu},
 \author[NANKAI]{Yu-Bin Liu},
 \author[ITP,PKU]{Jian-Ping Ma},
 \author[NANKAI]{Xiang-Fei Meng},
 \author[PKU]{Yan Shen},
 \author[ZHEJIANG]{and Jian-Bo~Zhang}
 \address[PKU]{School of Physics, Peking University\\
               Beijing, 100871, P.~R.~China}
 \address[IHEP]{Institute of High Energy Physics\\
                 Academia Sinica, P.~O.~Box 918\\
                 Beijing, 100039, P.~R.~China}
 \address[ITP]{Institute of Theoretical Physics\\
                 Academia Sinica, Beijing, 100080, P.~R.~China}
 \address[NANKAI]{Department of Physics, Nankai University\\
                  Tianjin, 300071, P.~R.~China}
 \address[ZHEJIANG]{Department of Physics, Zhejiang University\\
                  Hangzhou, 310027, P.~R.~China}
\thanks[fund]{This work is supported in part by the  National Natural
Science Foundation of China (NSFC) under grant No. 10421503, No.
10675005, No.10575107, No.10375031, No.10675101 and supported by the
Trans-century fund and the Key Grant Project of Chinese Ministry of
Education (No. 305001) and KJCX3-SYW-N2 (CAS).}

 \begin{abstract}
 We propose to study hadron-hadron scattering using lattice QCD in an asymmetric
 box which allows one to access more non-degenerate low-momentum modes for a given
 volume. The conventional L\"{u}scher's formula applicable in a symmetric box is
 modified accordingly. To illustrate the feasibility of this
 approach, pion-pion elastic scattering phase shifts in the $I=2$, $J=0$ channel
 are calculated within quenched approximation using improved gauge
 and Wilson fermion actions on anisotropic lattices in an asymmetric
 box. After the chiral and continuum extrapolation, we find that our
 quenched results for the scattering phase shifts
 in this channel are consistent with the experimental data
 when the three-momentum of the pion is below $300$MeV. Agreement is
 also found when compared with previous theoretical results from lattice
 and other means. Moreover, with the usage of asymmetric volume, we are able
 to compute the scattering
 phases in the low-momentum range (pion three momentum less than
 about $350$MeV in the center of mass frame)
 for over a dozen values of the pion three-momenta, much more than using
 the conventional symmetric box with comparable volume.
 \end{abstract}


 \end{frontmatter}

 \section{Introduction}
 Hadron-hadron scattering experiments have offered us
 enormous amount of information concerning the interaction among hadrons.
 In these experiments, scattering cross sections and phase shifts are obtained
 experimentally in various channels with definite quantum numbers.
 On the theoretical side, although Quantum Chromodynamics (QCD) has been recognized as the
 underlying theory of strong interaction, theoretical explanation
 of hadronic scattering processes at low
 energies remains a challenging problem due to non-perturbative features of
 the theory thereof. Lattice QCD (LQCD) is the only systematic, non-perturbative
 method of QCD which in principle can be applied  to calculate these low
 energy physical quantities from first principles using numerical Monte Carlo simulations.
 Calculation of hadron-hadron scattering phase shift is also a very important step to
 deepen our understanding of the strong interaction beyond single hadron spectrum.

 Lattice calculation of hadron scattering processes relies on
 a finite size method proposed by
 M.~L\"{u}scher~\cite{luscher86:finitea,luscher86:finiteb,luscher90:finite,luscher91:finitea,luscher91:finiteb}
 in which two particle elastic scattering
 phase shifts (in the infinite volume) are directly related to the
 energy levels of the two particles in a
 finite cubic box. The latter can in principle be extracted in lattice simulations.
 Using this technique, the scattering length and the scattering phase shifts
 for pion-pion scattering in the $I=2$, $J=0$ channel have been
 studied~\cite{gupta93:scat,fukugita95:scat,jlqcd99:scat,JLQCD02:pipi_length,chuan02:pipiI2,juge04:pipi_length,chuan04:pipi,ishizuka05:pipi_length,%
 CPPACS03:pipi_phase,CPPACS04:pipi_phase_unquench,savage06:pipi}
 in both quenched and unquenched lattice QCD.
 There also exist lattice calculations on other hadronic scattering processes
 using various lattice actions.

 The lattice results on the pion-pion scattering phases can be compared
 with the experimental data and results from other theoretical
 methods and impressive agreements were
 seen~\cite{CPPACS03:pipi_phase,CPPACS04:pipi_phase_unquench,savage06:pipi}.
 However, since lattice results were obtained in a finite volume,
 the number of low-momentum modes accessible for lattice simulation
 was limited. Part of the reason is that, in all previous lattice studies,
 hadron scattering phase shifts  were calculated in a cubic box which
 has the same physical extension in all three spatial directions. In this
 scenario, many low momentum modes are degenerate in energy such as modes
 $(1,0,0)$, $(0,1,0)$ and $(0,0,1)$ since they are related to one another
 by cubic symmetry. As a result, one can only access very few
 low-momentum modes in the lattice calculation in a cubic box.
 If one is interested in the scattering phases at more values of the scattering momenta,
 larger physical volumes are required which makes the lattice simulation very
 costly. In this paper, we propose to use asymmetric boxes to
 study hadron-hadron scattering on the lattice. We test this
 idea in a quenched study on the pion-pion scattering in the
 $I=2$, $J=0$ channel. If we denote the three-momentum of each pion in the center of
 mass frame by $\bar{k}$, within the range $0<\bar{k}^2<0.1$GeV$^2$, we
 are able to obtain scattering phase shifts at more than $12$ different
 values of $\bar{k}^2$ while in calculations with comparable cubic volumes
 this number is restricted to only a few.

 This paper is organized as follows. In Section 2, we briefly review
 the theoretical formalism for the computation of the phase shift in an asymmetric box,
 extending the finite size technique suggested by L\"uscher.
 The corresponding formulae are modified to the case of asymmetric
 volume. Possible mixing with the $J=2$ channel is discussed.
 In Section 3, some simulation details are given.
 Our results for the scattering phases, after chiral and continuum extrapolations,
 are then compared with known results from previous lattice calculations,
 chiral perturbation theory, dispersion relations and experimental data.
 Reasonable agreements are found and the advantage of using the asymmetric box is addressed.
 In last section, we will summarize this work and give some conclusions
 and outlooks.

 \section{L\"uscher's formulae extended to an asymmetric box}

 Consider a cubic box with size $L \times L \times L$ and periodic boundary condition.
 In such a box, the three momentum
 of a single pion is quantized as: $\textbf{k}=(2\pi/L)\textbf{n}$
 with $\textbf{n}=(n_1,n_2,n_3)\in \mathbb{Z}^3$, where $\mathbb{Z}$ represents the
 set of all integers. In this paper, we are interested in two-pion systems.
 Taking the center of mass reference frame of the two pions, we define
 $\bar{\textbf{k}}^2$ of the pion pair in a box as:
 \begin{equation}
 \label{eq:kbar_def}
 E_{\pi\pi}=2\sqrt{m_{\pi}^2+\bar{\textbf{k}}^2}
 \end{equation}
 where $E_{\pi\pi}$ is the {\em exact} energy of a two-pion system
 with the two pions having three momentum $\textbf{k}$ and $-\textbf{k}$
 respectively in the center of mass frame.
 We further define $q^2$ via:
 \begin{equation}
 \label{eq:q_def}
 \bar{\textbf{k}}^2=\frac{4\pi^2}{L^2}q^2
 \end{equation}
 Note that, because of the interaction between the two
 pions in the box, the value of $q^2$ is in general not equal to
 $\textbf{n}^2$ with $\textbf{n}\in \mathbb{Z}^3$.
 In fact, in the $I=2$, $J=0$ channel, the interaction
 between the two pions is repulsive,
 making the value of $q^2$ larger than the corresponding $\textbf{n}^2$.

 According to L\"{u}scher's method~\cite{luscher91:finitea},
 two-pion $s$-wave elastic scattering phase shift
 can be obtained from the following formula:
 \footnote{This assumes that the contribution from
 higher angular momentum modes are negligible. In the
 cubic case, the leading contamination is from $l=4$.}
 \begin{equation}
 \label{eq:original_luscher}
 \cot{\delta (\bar{k})}=\frac{\mathcal{Z}_{00}(1,q^2)}{\pi^{3/2}q}
 \;,
 \end{equation}
 where the so-called zeta function $\mathcal{Z}_{lm}$ is given by:
 \begin{equation}
 \mathcal{Z}_{lm}(s,q^2)=\sum_{\textbf{n}}
 \frac{\mathcal{Y}_{lm}(\textbf{n})}{(\textbf{n}^2-q^2)^s}
 \;.
 \end{equation}
 In this definition,
 $\mathcal{Y}_{lm}(\textbf{r})=r^l\mathit{Y}_{lm}(\Omega_\textbf{r})$,
 with $Y_{lm}(\Omega_\textbf{r})$ being the usual spherical harmonics.
 When the physical volume is large enough, L\"uscher's
 formula~(\ref{eq:original_luscher})
 can be expanded as powers of $1/L$. The resulting formula
 then relates the energy level of two
 hadron in a finite box to the hadron elastic scattering length in
 the infinite volume. This provides a very convenient way of
 computing scattering lengths on the lattice. The formula reads:
 \begin{equation}
 E_{\pi\pi}-2m_{\pi}=-\frac{4\pi a_0}{m_\pi
 L^3}\left[1+c_1\left(\frac{a_0}{L}\right)
 +c_2\left(\frac{a_0}{L}\right)^2\right]+O(L^{-6})
 \end{equation}
 where the coefficients $c_1=-2.837297$ and $c_2=6.375183$,
 $a_0$ is the $\pi\pi$ elastic scattering length.

 As explained in the introduction of this paper, many
 low momentum modes in a cubic box are degenerate in energy due to cubic symmetry.
 Therefore, to compute the scattering phases at more values
 of the scattering momenta, one usually has to use larger
 cubic volumes. This makes the lattice calculation more costly.
 In this paper, we propose to use an asymmetric volume which
 provides more non-degenerate low-momentum modes with a relatively small volume.
 L\"{u}scher's original formula~(\ref{eq:original_luscher}) is only valid in a cubic box.
 To utilize similar finite volume techniques, we must generalize
 Eq.~(\ref{eq:original_luscher}) to the case of an asymmetric box.
 This has been accomplished in~\cite{chuan04:asymmetric,chuan04:asymmetric_long}.

 In an asymmetrical box with lattice size $L \times \eta_2 L \times \eta_3 L$,
 the momentum of a single pion is quantized as:
 $\textbf{k}=(2\pi/L)\tilde{\textbf{n}}$ with
 $\tilde{\textbf{n}}\equiv(n_{1},n_{2}/\eta_2,n_{3}/\eta_3)$ and
 $\textbf{n}=(n_1,n_2,n_3)\in\mathbb{Z}^3$.
 Quantities $\bar{\textbf{k}}$ and
 $q^2$ are still defined according to Eq.~(\ref{eq:kbar_def}) and Eq.~(\ref{eq:q_def}).

 The symmetry group of the asymmetric box
 depends on the shape of the volume we take in our lattice calculation.
 For definiteness, we choose $\eta_2=1$ and $\eta_3=2$ in this study and the
 corresponding basic symmetry group is $D_4$ which has 4 one-dimensional
 representations: $A_1$, $A_2$, $B_1$, $B_2$ and a two-dimensional
 irreducible representation $E$. Rotational symmetry is broken and
 the corresponding representations for the rotational
 group with definite angular momentum quantum numbers
 are decomposed accordingly:
 \begin{equation}
 \textbf{0}=A^+_1\;,
 \;\;
 \textbf{1}=A^-_2+E^-\;,
 \;\;
 \textbf{2}=A^+_1+B^+_1
 +B^+_2+E^+\;, \;\;\cdots\;.
 \end{equation}

 The formula for the scattering phase shifts is now modified to:
 \footnote{Again, we omit higher angular momentum contributions. In
 this case, the leading contamination is from $l=2$.}
 \begin{equation}
 \label{eq:luescher_modified}
 \cot{\delta
 (\bar{k})}= m_{00}(q)\equiv \frac{\mathcal{Z}_{00}(1,q^2;\eta_{2},\eta_{3})}{\pi^{3/2}\eta_{2}\eta_{3}q}
 \end{equation}
 with the modified zeta function $\mathcal{Z}_{lm}$ defined as:
 \begin{equation}
 \mathcal{Z}_{lm}(s,q^2;\eta_{2},\eta_{3})=
 \sum_{\textbf{n}}\frac{\mathcal{Y}_{lm}(\tilde{\textbf{n}})}{(\tilde{\textbf{n}}^2-q^2)^s}
 \end{equation}
 The formula for scattering length is also changed accordingly:
 \begin{equation}
 \label{eq:a0_asymmetric}
 E_{\pi\pi}-2m_{\pi}=-\frac{4\pi a_0}{\eta_2\eta_3m_\pi
 L^3}\left[1+c_1(\eta_2,\eta_3)\left(\frac{a_0}{L}\right)
 +c_2(\eta_2,\eta_3)\left(\frac{a_0}{L}\right)^2\right]+O(L^{-6})
 \;.
 \end{equation}
 where the coefficients $c_1(\eta_2,\eta_3)$ and $c_2(\eta_2,\eta_3)$ can be
 computed once $\eta_2$ and $\eta_3$ are given~\cite{chuan04:asymmetric}.
 For the case $\eta_2=1$ and $\eta_3=2$, which is the situation studied in
 this paper, the two coefficients are found to be:
 \begin{equation}
 c_1(1,2)=-1.805872\;,
 \;\;
 c_2(1,2)=1.664979
 \end{equation}
 Therefore, just as in the cubic case, once the two-pion energy
 level $E_{\pi\pi}$ is obtained in Monte Carlo simulations,
 the corresponding phase shift $\delta$ can be obtained via
 modified L\"uscher's formula~(\ref{eq:luescher_modified}).

 It is noted that the representation $A_1$ appears in both the $J=0$ and
 $J=2$ channel. Therefore, in an asymmetric box, $s$-wave and
 $d$-wave scattering mix with each other. This is to be compared
 with the cubic case where the lowest mixture to $s$-wave is
 from $l=4$ sector. Assuming the $d$-wave scattering phases are
 small, one can estimate its effect on the $s$-wave phase shift as
 follows~\cite{chuan04:asymmetric_long}:
 \begin{equation}
 n\pi-\delta_0(q)\simeq \phi(q)
 +\sigma_2(q)\tan\delta_2(q)\;,
 \end{equation}
 where the angle $\phi(q)$ is defined via:
 $-\tan\phi(q)=1/m_{00}(q)$. The function $\sigma_2(q)$ for
 $D_4$ symmetry is given by:
 \begin{equation}
 \sigma_2(q)= {m^2_{02}(q)\over 1+m^2_{00}(q)}\;,
 \end{equation}
 which quantifies the effect due to $d$-wave mixing.
 \footnote{The explicit formula for the function $m_{02}(q)$ can
 be found in Ref.~\cite{chuan04:asymmetric_long}.}
 On general grounds, one expects the mixing due to higher
 angular momentum to be small in the low-momentum region.
 To estimate its effect in the case of pion-pion scattering,
 we take experimental values for the $d$-wave scattering phases
 presented in Ref.~\cite{Hoogland:1974cv}.
 We have checked this correction to our $s$-wave phase shifts and it is found that
 these corrections are only of about $1-2$\%, much smaller
 than our typical error bars for the phase shifts. Therefore,
 in what follows, we simply neglect the effects of the $d$-wave contaminations.

 \section{Simulation Details}

 To test our idea of using the asymmetric box on hadron-hadron
 scattering, we perform a quenched study on the pion-pion
 scattering phase shift in the $I=2$, $J=0$ channel. In this
 section, we will briefly introduce our numerical results.
 \begin{table}
 \caption{Simulation parameters used in this work:\label{tab:parameter}}
 \begin{tabular}{|c||c|c|c|c|c|c|c|}
 \hline
 $\beta$ & $u_s$ & $\nu$ & Lattice & $a_s(GeV^{-1})$ & Number of Confs.& $\kappa_{max}$\\
 \hline
 2.080 & 0.7735 & 0.94 & $8^2\times 16\times 40$ & 1.5677 & 464 & 0.0598\\
 \hline
 2.215 & 0.7852 & 0.95 & $9^2\times 18\times  48$ & 1.3926 & 425 & 0.0602\\
 \hline
 2.492 & 0.8063 & 0.93 & $12^2\times 24\times 64$ & 1.0459 & 105 & 0.0606\\
 \hline
 \end{tabular}
 \end{table}

 \subsection{Lattice actions and simulation parameters}

 The gauge action used in this study is the tadpole improved gluonic
 action on anisotropic lattices~\cite{colin97,colin99}:
 \begin{displaymath}
 \label{eq:gauge_action}
 S=-\beta\sum_{i>j}[\frac{5}{9}\frac{TrP_{ij}}{\xi
 u^4_s}-\frac{1}{36}\frac{TrR_{ij}}{\xi
 u^6_s}-\frac{1}{36}\frac{TrR_{ji}}{\xi u^6_s}]
 \end{displaymath}
 \begin{equation}
 -\beta\sum_i[\frac{4}{9}\frac{\xi TrP_{01}}{
 u^2_s}-\frac{1}{36}\frac{\xi TrR_{i0}}{u^4_s}]
 \end{equation}
 where $P_{ij}$ is the usual spatial plaquette variables and $R_{ij}$ is the
 $2\times 1$ spatial Wilson loop on the lattice. The parameter $u_s$,
 which we take to be the $4$-th root of the average spatial plaquette
 value, incorporates the so-called tadpole improvement~\cite{lepage93:tadpole} and $\xi$
 designates the (bare) aspect ratio of the anisotropic lattice, defined as the ratio
 between two spacings $a_s/a_t$. With the tadpole improvement,
 experiences show that the renormalization effects are small for this
 parameter. Thus, we have not distinguish the renormalized anisotropy
 and the bare one. The anisotropic and improvement property of the
 lattice action makes the calculation of  heavier hadronic objects
 on coarser lattice possible. The parameter
 $\beta$ is related to the bare gauge coupling which controls the spatial
 lattice spacing $a_s$ in physical units.
 This type of improved gauge action on anisotropic
 lattices have been extensively used in lattice calculations on
 glueballs~\cite{colin99,chuan01:gluea,chuan01:glueb,chuan01:india,chen_liu06:glueball}.

 The fermion action used in this calculation is the tadpole improved clover Wilson action on
 anisotropic lattice~\cite{klassen98:wilson_quark,klassen99:aniso_wilson,chuan01:tune} whose
 fermion matrix reads:
 $\mathcal{M}_{xy}=\delta_{xy}\sigma+\mathcal{A}_{xy}$ with $\mathcal{A}$
 given by:
 \begin{displaymath}
 \mathcal{A}_{xy}=\delta_{xy}[1/(2\kappa_{max})+\rho_{t}
 \sum^3_{i=1}\sigma_{0i}\mathcal{F}_{0i}+\rho_s(\sigma_{12}\mathcal{F}_{12}
 +\sigma_{23}\mathcal{F}_{23}+\sigma_{31}\mathcal{F}_{31})]
 \end{displaymath}
 \begin{equation}
 -\sum_{\mu}\eta_{\mu}[(1-\gamma_{\mu})U_{\mu}(x)\delta_{x+\mu,y}
 +(1+\gamma_{\mu})U^+_{\mu}(x-\mu)\delta_{x-\mu,y}]
 \end{equation}
 where the coefficients are given by:
 \begin{displaymath}
 \eta_i=\nu /(2u_s),
 \eta_0=\xi/2,\sigma=1/(2\kappa)-1/(2\kappa_{max}),
 \end{displaymath}
 \begin{equation}
 \rho_t=c_{SW}(1+\xi)/(4u^2_s),\rho_s=c_{SW}/(2u^4_s).
 \end{equation}\\
 In this notation, the fermion propagators with different quark masses could
 be solved at the same time using the so-called Multi-mass Minimal Residual ($M^3R$)
 algorithm~\cite{frommer95:multimass,glaessner96:multimass,beat96:multimass}.
 The bare velocity of light parameter $\nu$ is tuned non-perturbatively using the single
 pion dispersion relations~\cite{chuan01:tune,chuan06:tune_v}. The parameters $\kappa_{max}$
 is the largest one among all $\kappa$ parameters which corresponds
 to the lightest valence quark mass.  The asymmetrical ratio $\xi$ is always
 fixed at $\xi=5$. Other parameters in our simulation are tabulated
 in Table~\ref{tab:parameter}.

 Quenched configurations are generated using the pure gauge
 action~(\ref{eq:gauge_action}) with three lattice sizes,
 $8^2\times16 \times 40$, $9^2\times 18 \times 48$ and
 $12^2\times 24 \times 64$, corresponding to
 $\beta=2.080$, $2.215$ and $2.492$,
 respectively. The values of $\beta$ are chosen such that the
 physical volumes for these three lattices remain the same.
 The correspondence of $\beta$ and the spatial lattice
 spacing $a_s$ has been obtained in Ref.~\cite{chuan06:tune_beta}.
 The physical volume of our lattices is about $5.5fm^3$ which is large
 enough to bring the finite volume errors under control. For each set
 of parameters, several hundreds of de-correlated gauge
 configurations are utilized to measure physical quantities.

 \subsection{Hadronic operators and the extraction of one and two pion energies}
 To obtain energy levels for the single and two-pion systems on the
 lattice, we have to construct appropriate correlation functions
 using the corresponding hadronic operators.
 In this paper, single and two pion operators are constructed using local quark fields.
 For the single pion operators, we use:
 \begin{displaymath}
 \pi^+(\textbf{x},t)=-\bar{d}(\textbf{x},t)r_5u(\textbf{x},t),
 \pi^-(\textbf{x},t)=\bar{u}(\textbf{x},t)r_5d(\textbf{x},t)
 \end{displaymath}
 \begin{equation}
 \pi^0(\textbf{x},t)=\frac{1}{\sqrt{2}}(\bar{u}(\textbf{x},t)r_5d(\textbf{x},t)
 -\bar{d}(\textbf{x},t)r_5u(\textbf{x},t))
 \end{equation}
 where $u(\textbf{x},t)$ and $d(\textbf{x},t)$ are the basic local
 quark field operators for the up and down quark, respectively.  In
 this study, the up and down quarks are taken to be degenerate
 in mass so that isospin is a good symmetry.
 The operator which creates a single pion with non-zero three momentum
 $\textbf{k}$ from the vacuum is obtained by Fourier transform:
 \begin{equation}
 \pi^a_{\textbf{k}}(t)=\frac{1}{V_3}\sum_{\textbf{x}}
 \pi^a(\textbf{x},t)e^{-i\textbf{k}\cdot\textbf{x}}
 \end{equation}
 where the flavor index $a$ of pions take values $a=+,-,0$ and
 $V_3$ is the three volume of the lattice. By calculating
 correlation functions of single pion operators defined above,
 one can obtain the single pion energy at vanishing and
 non-vanishing momenta.

 The $s$-wave two-pion operators in the $I=2$ channel are defined as:
 \begin{equation}
 \label{eq:twopion_def}
 \mathcal{O}_n(t)=\sum_{R}\pi^+_{R(\textbf{k}_n)}(t)\pi^+_{R(-\textbf{k}_n)}(t+1)
 \;,
 \end{equation}
 where $n$ labels a particular mode with three-momentum
 $\textbf{k}_n$; $R(\textbf{k}_n)$ is the rotated three-momentum
 which is obtained from $\textbf{k}_n$ by applying a symmetry
 operation $R\in D_4$, an element of the corresponding point group.
 Therefore, the summation of $R$ in Eq.~(\ref{eq:twopion_def})
 guarantees that the operator thus constructed falls
 into the $A^+_1$ representation of the symmetry group whose continuum counterpart is
 $s$-wave for the rotational group, if the
 contaminations from  the $l\geq 2$ sectors are negligible.
 \begin{table}
 \caption{The representative momentum of every mode. Mode 2, 4 and 5
 have accidental degenerate modes which are left out in the construction
 of two pion operators.\label{tab:modes}}
 \begin{tabular}{|c||c|c|c|c|c|c|c|}
 \hline
 Serial Number & 0 & 1 & 2 & 3 & 4 & 5 & 6 \\
 \hline
 Mode $\tilde{\textbf{n}}$ & (0,0,0) & (0,0,1/2) & (1,0,0) &
 (1,0,1/2) & (1,1,0) & (1,1,1/2) & (1,1,1)\\
 \hline
 Degenerate Mode $\tilde{\textbf{n}}$ &&& (0,0,1) & & (1,0,1) & (0,0,3/2)& \\
 \hline
 \end{tabular}
 \end{table}

 In order to obtain the two-pion energies, which is directly related
 to the scattering phases we want to compute, we measure the
 correlation matrix among different non-degenerate two-pion modes,
 using the two-pion operators defined in Eq.~(\ref{eq:twopion_def}):
 \begin{equation}
 \label{eq:correlation_matrix_def}
 \mathcal{C}_{mn}(t)=\langle\mathcal{O}^\dagger_m(t)\mathcal{O}_n(t_s)\rangle \;.
 \end{equation}
 We then follow L\"{u}scher
 and Wolff's suggestion and constructed a new correlation matrix:
 \begin{equation}
 \Omega(t,t_0)=\mathcal{C}(t_0)^{-\frac{1}{2}}\mathcal{C}(t)\mathcal{C}(t_0)^{-\frac{1}{2}}
 \end{equation}
 where $t_0$ is some suitable reference time. The eigenvalue
 $\lambda_i(t)$ of this new matrix $\Omega$ is:
 \begin{equation}
 \lambda_i(t,t_0)\propto e^{-E_i(t-t_0)}
 \end{equation}
 It can be shown that~\cite{luscher90:finite} this eigenvalue avoids
 $\mathcal{O}(e^{-\Delta Et})$
 errors and the energy eigenvalues could be extracted by a single
 exponential in $t$.
 We choose seven non-degenerate momentum modes to construct the
 correlation function matrix~(\ref{eq:correlation_matrix_def}).
 The representative momentum of each
 non-degenerate momentum mode is tabulated in Table~\ref{tab:modes}. Note that some modes
 might become accidentally degenerate in the continuum limit when rotational invariance is completely
 restored. That is to say, two modes are degenerate in energy in the continuum limit
 but they are not related to one another by any $D_4$ transformation.
 In Table~\ref{tab:modes} we also list these accidental degenerate modes.
 Although for finite lattice spacings,
 scaling violations will lift these degeneracies, the almost
 degenerate modes might make the diagonalization procedure unstable.
 Therefore, in our study of the two-pion correlation matrix,
 the accidental degenerate modes are left out in the construction
 of the two-pion operators.

\begin{figure}[htb]
 \begin{center}
 \subfigure{\includegraphics[width=0.52\textwidth]{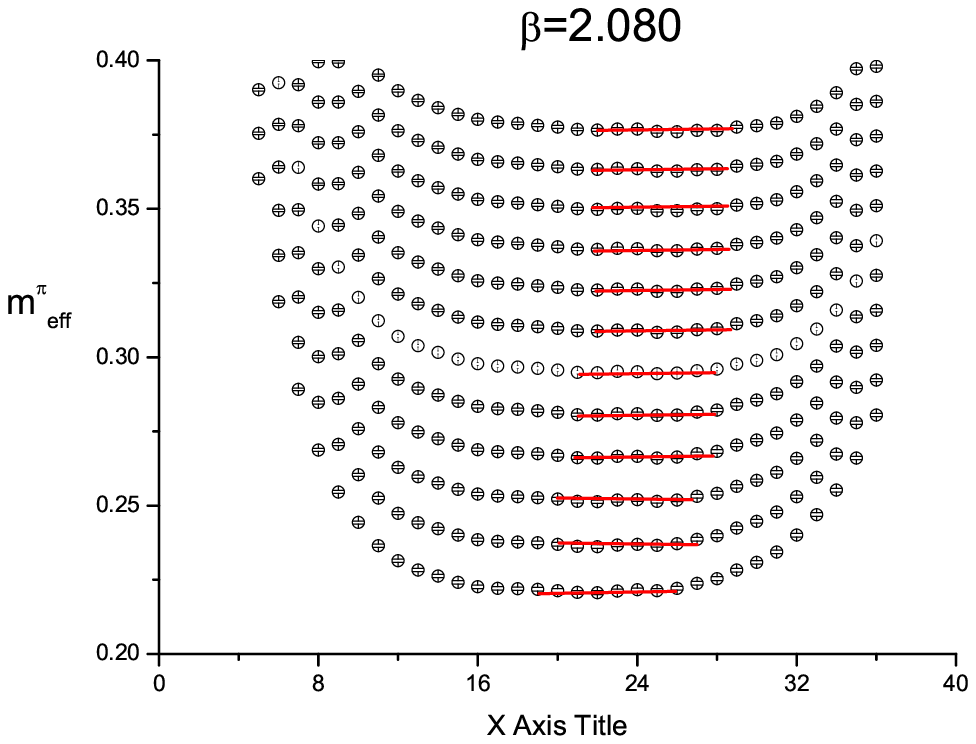}}
 \subfigure{\includegraphics[width=0.52\textwidth]{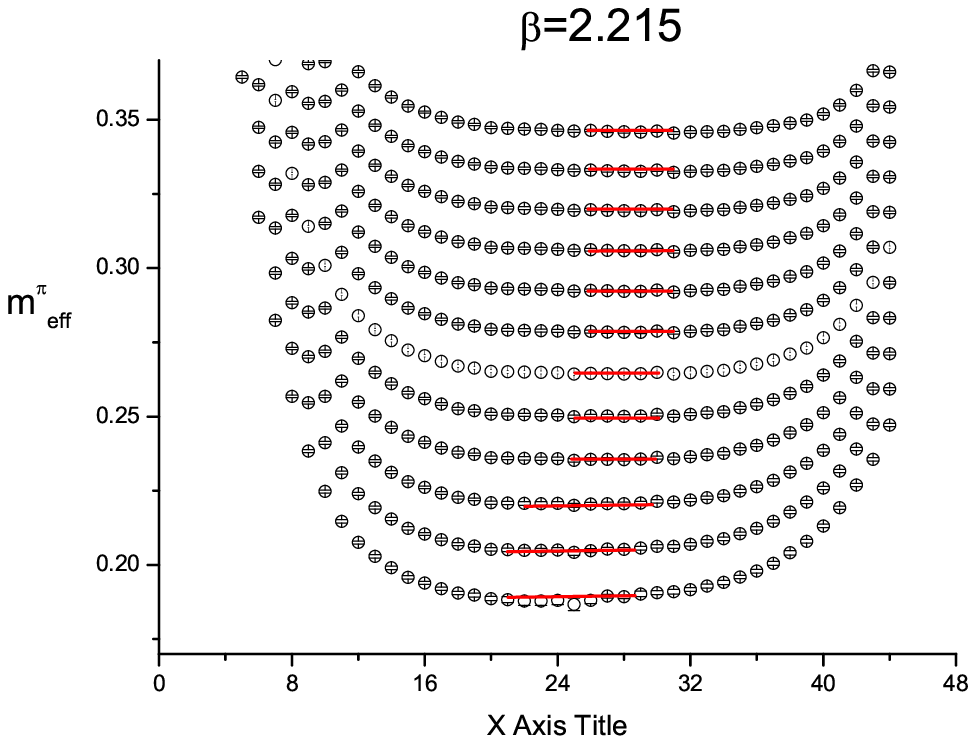}}
 \subfigure{\includegraphics[width=0.52\textwidth]{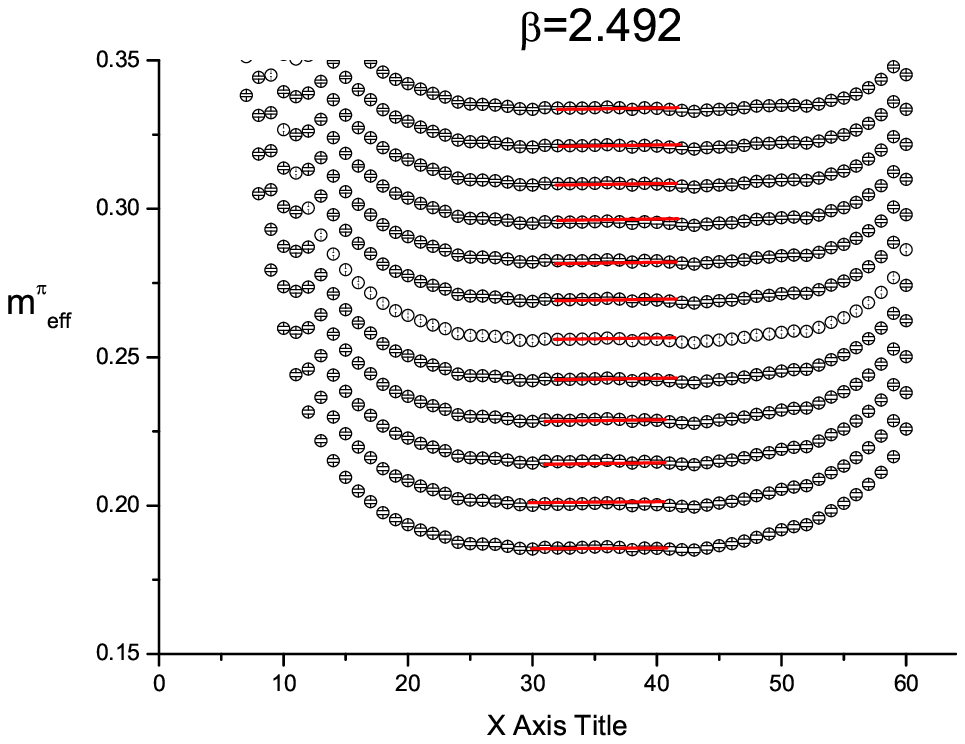}}
 \end{center}
 \caption{The single pion effective mass plateaus are shown. The
 horizontal line segments in the figure represent the fitting ranges
 of the plateaus.\label{fig:single_pion}}
 \end{figure}

 The single pion correlations at zero spacial momentum are
 constructed from the wall source quark propagators. Effective mass
 functions are then used to extract the single pion mass values. The
 mass plateaus are determined automatically by requiring the minimal
 of $\chi^2$ per degree of freedom.
 All effective mass plateaus are plotted in Fig.~\ref{fig:single_pion}.
 The horizontal line segments in these figures represent
 the ranges of the plateaus from which the pion masses are extracted.
 The errors for the data points are obtained from a standard jack-knife analysis.

\begin{figure}[htb]
 \begin{center}
 \subfigure{\includegraphics[width=0.45\textwidth]{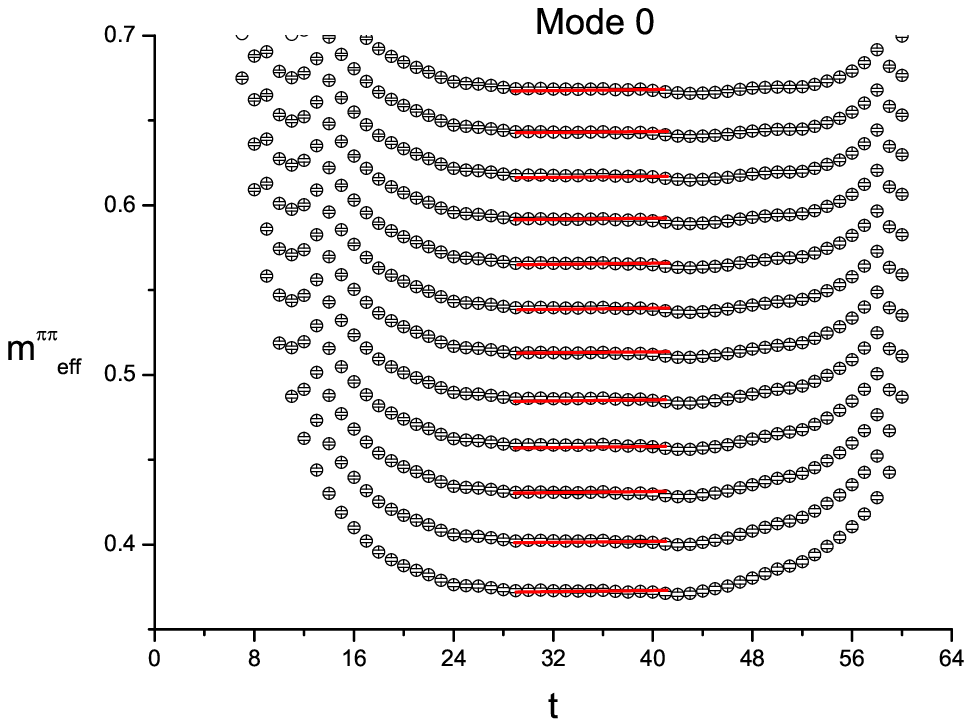}}
 \subfigure{\includegraphics[width=0.45\textwidth]{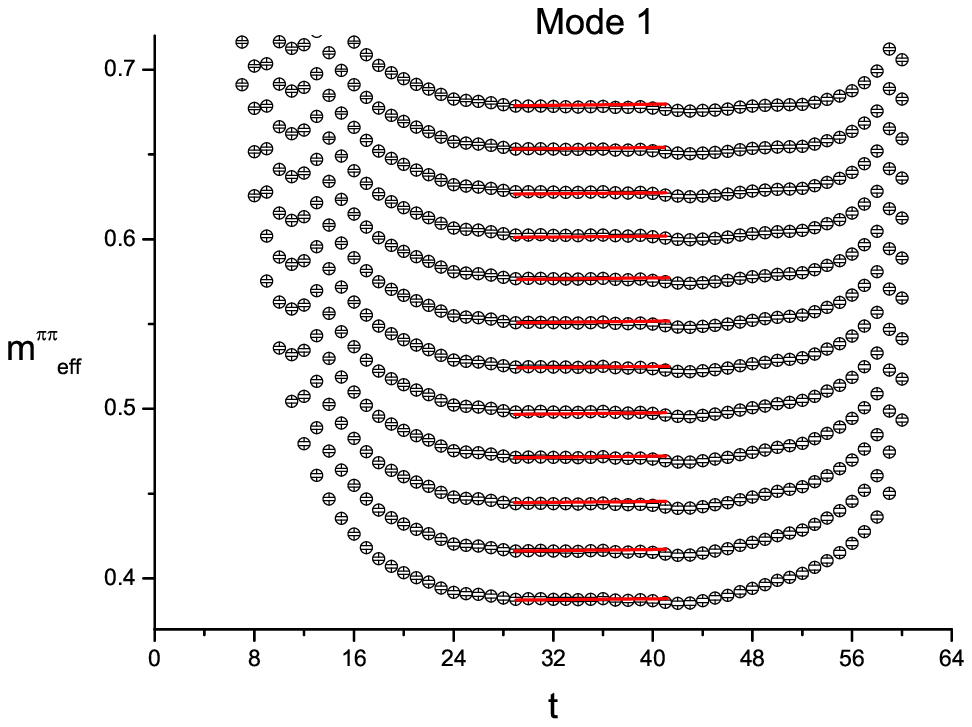}}
 \subfigure{\includegraphics[width=0.45\textwidth]{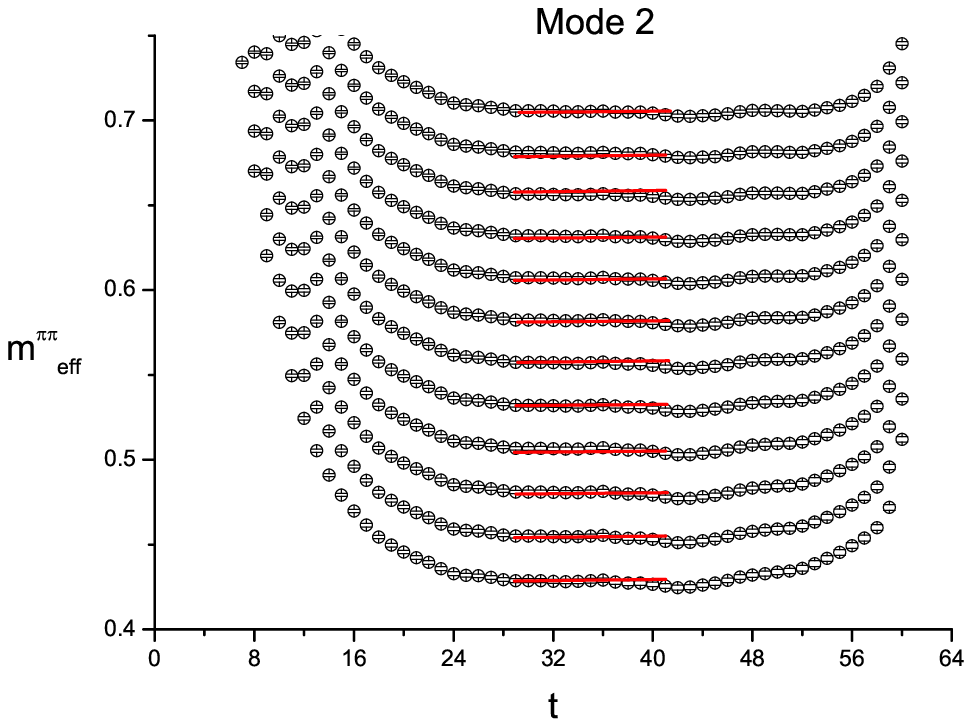}}
 \subfigure{\includegraphics[width=0.45\textwidth]{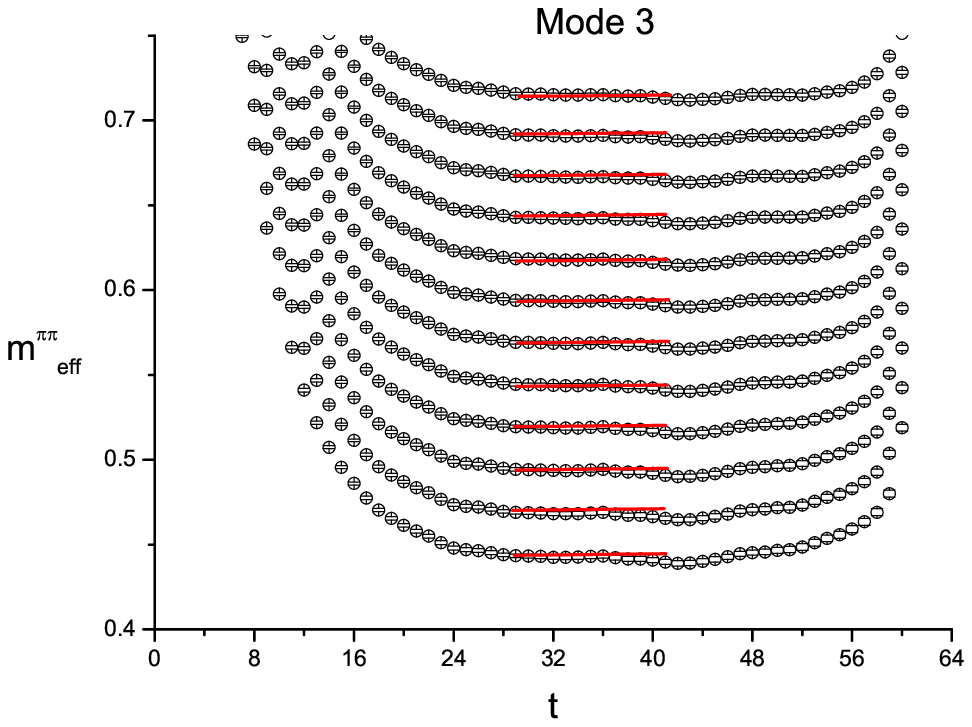}}
 \subfigure{\includegraphics[width=0.45\textwidth]{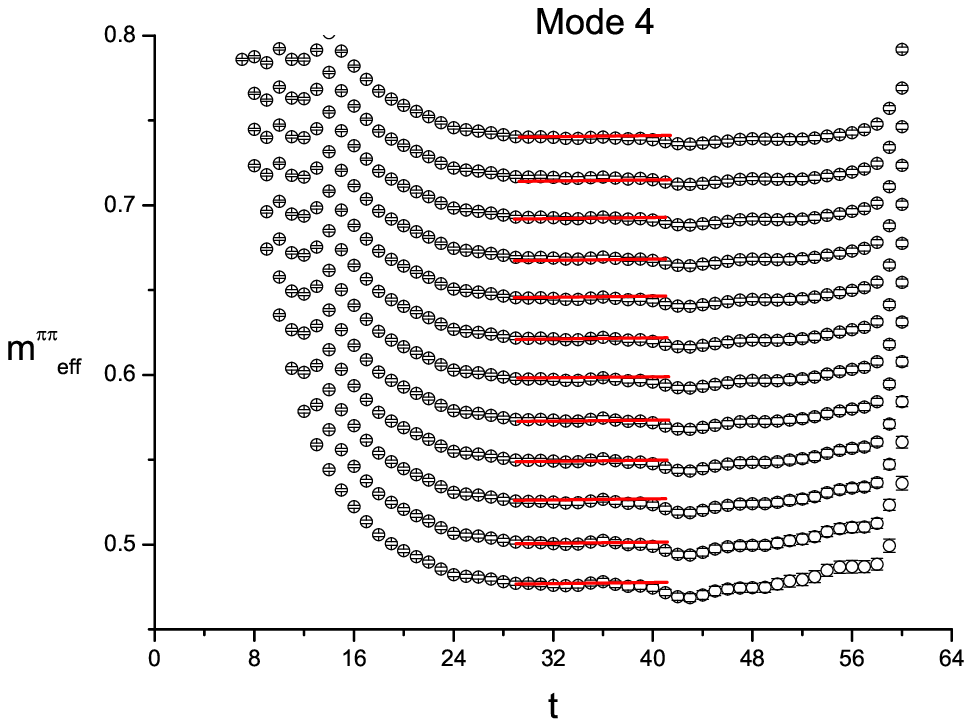}}
 \subfigure{\includegraphics[width=0.45\textwidth]{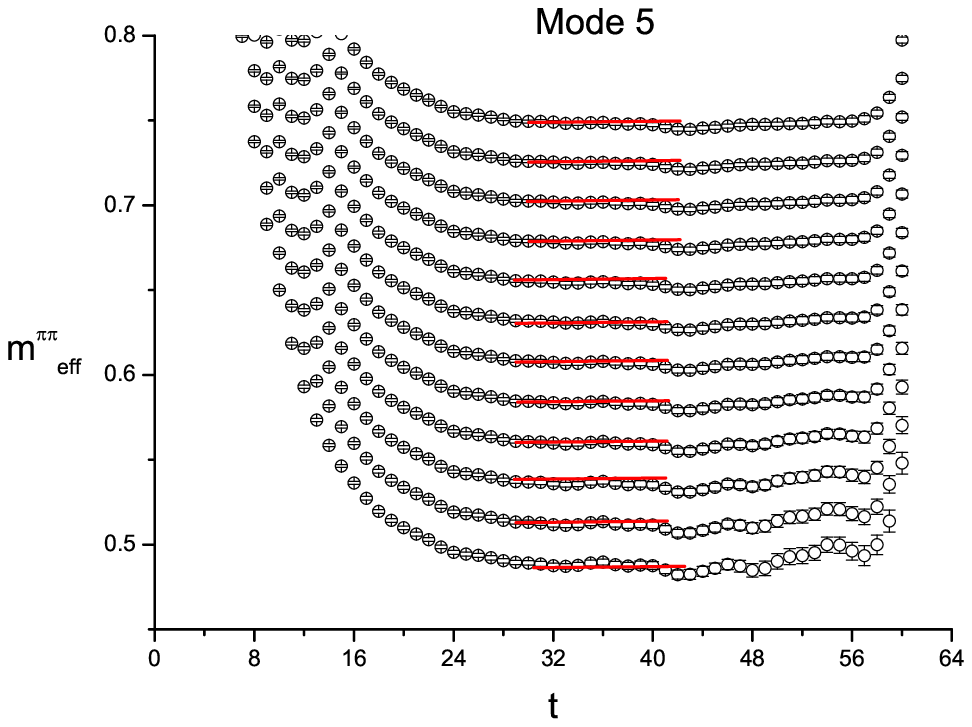}}
 \end{center}
 \caption{Effective mass plateaus of two-pion system at $\beta=2.492$
 after the diagonalization procedure. The horizontal line segments in
 the figure represent the fitting ranges of the plateaus. Results for
 other values of $\beta$ are similar.\label{fig:twopion_energy2492}}
 \end{figure}

 Similar analysis is performed for the two-pion correlation matrix.
 It is verified that  the symmetric off-diagonal matrix elements,
 $\mathcal{C}_{ij}$ and $\mathcal{C}_{ji}$ ($i \neq j$) of the matrix, are almost
 equal to each other in all cases except for some high modes at
 large time slices. Therefore, in our analysis, $\mathcal{C}_{ij}$ and
 $\mathcal{C}_{ji}$ ($i\neq j$) are simply averaged to construct a
 symmetric positive-definite matrix.
 In the diagonalization procedure, we set $t_0=3$ as the reference time.
 The effective energies of two-pion energy levels are defined as
 $E_{\pi\pi}(i,t)=\ln(\frac{\lambda_i(t)}{\lambda_i(t+1)})$ where the
 index $i$ represents the $i$-th eigenmode and the energy plateaus
 $E_{\pi\pi}$ are found accordingly. In
 Fig.~\ref{fig:twopion_energy2492}, the effective mass plateaus at $\beta=2.492$
 for various modes are shown. Data for other values of $\beta$ are
 similar. The corresponding values for
 $\bar{\textbf{k}}$ can thus be obtained from Eq.~(\ref{eq:kbar_def})
 and the values for $\delta$ may be computed by Eq.~(\ref{eq:luescher_modified})
 for each set of bare parameters.

 \subsection{Results for the scattering length}

 It is found that the relative three momentum
 $|\bar{\textbf{k}}|$ for Mode 0 is far smaller than the corresponding
 single-pion mass values. Therefore, $\delta E=E_{\pi\pi}-2m_{\pi}$ can be
 obtained from Mode 0 ignoring the small three momentum effect.
 It is then easy to obtain the scattering length $a_0$ in this
 channel using formula~(\ref{eq:a0_asymmetric}) for every valence
 quark mass and $\beta$ value. These results are then used to perform the
 chiral extrapolation.

\begin{figure}[htb]
 \begin{center}
 \subfigure{\includegraphics[width=0.48\textwidth]{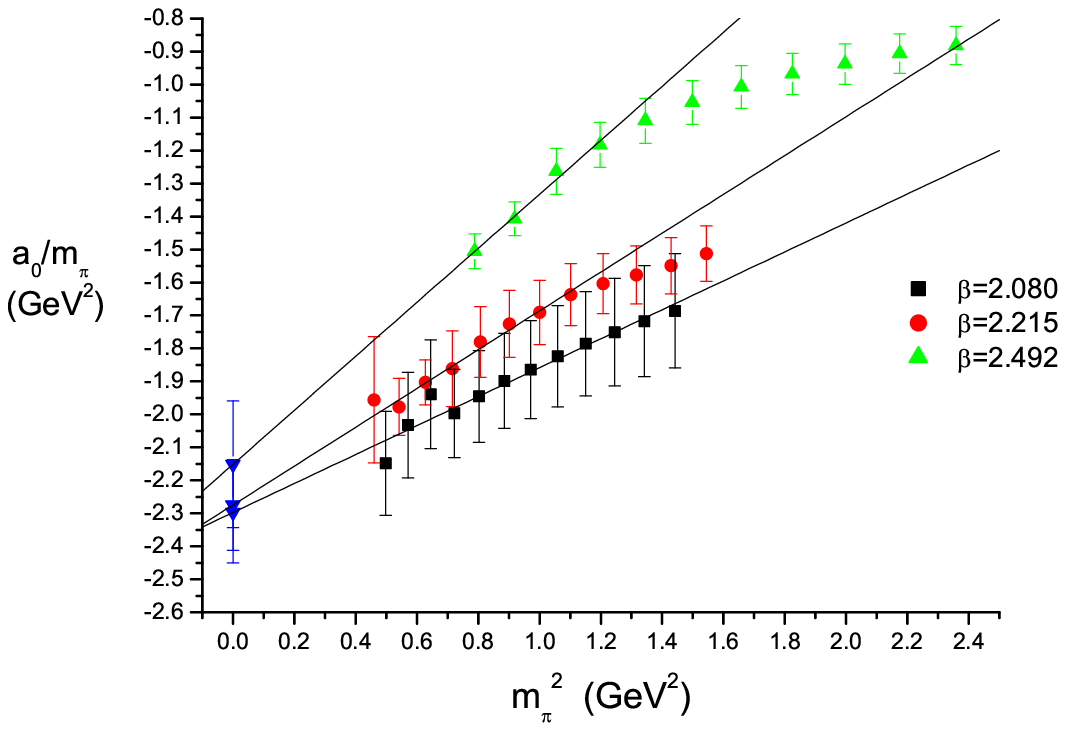}}
 \subfigure{\includegraphics[width=0.48\textwidth]{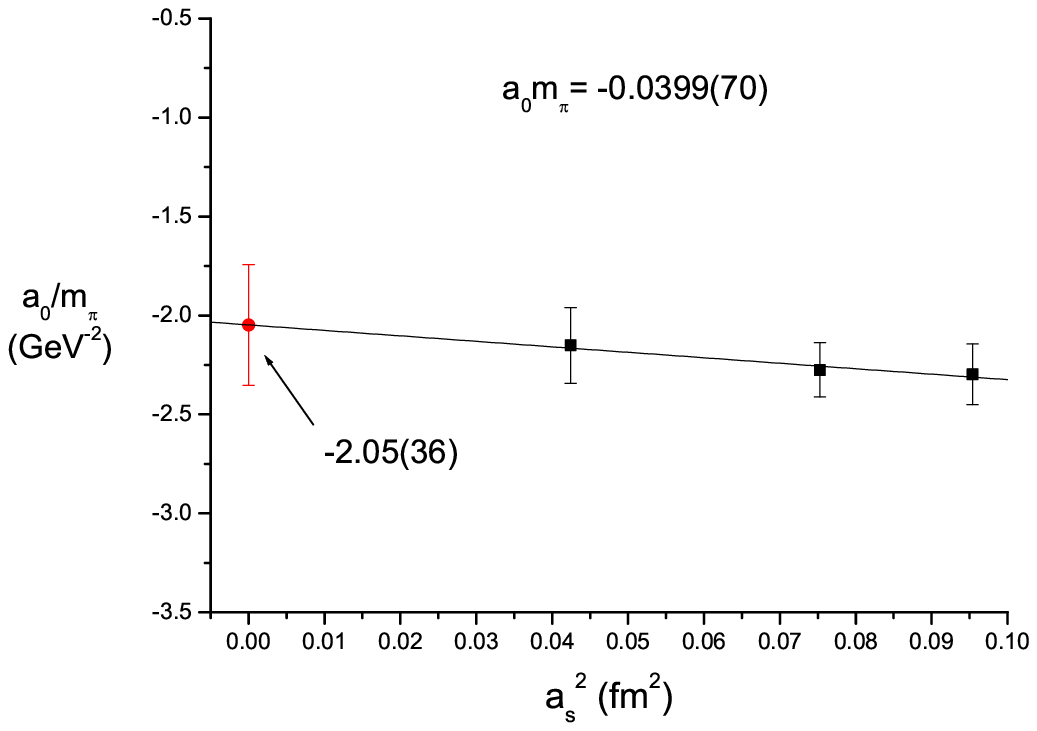}}
 \end{center}
 \caption{Left panel: chiral extrapolation of the quantity
 $\frac{a_0}{m_{\pi}}$ using Scheme 1 as described in the paper.
 Right panel:  the corresponding continuum limit extrapolation of the
 quantity $\frac{a_0}{m_{\pi}}$. The extrapolated result is indicated
 by a solid circle near $a_s=0$.\label{fig:delta_a0_linear}}
 \end{figure}

\begin{figure}[htb]
 \begin{center}
 \subfigure{\includegraphics[width=0.48\textwidth]{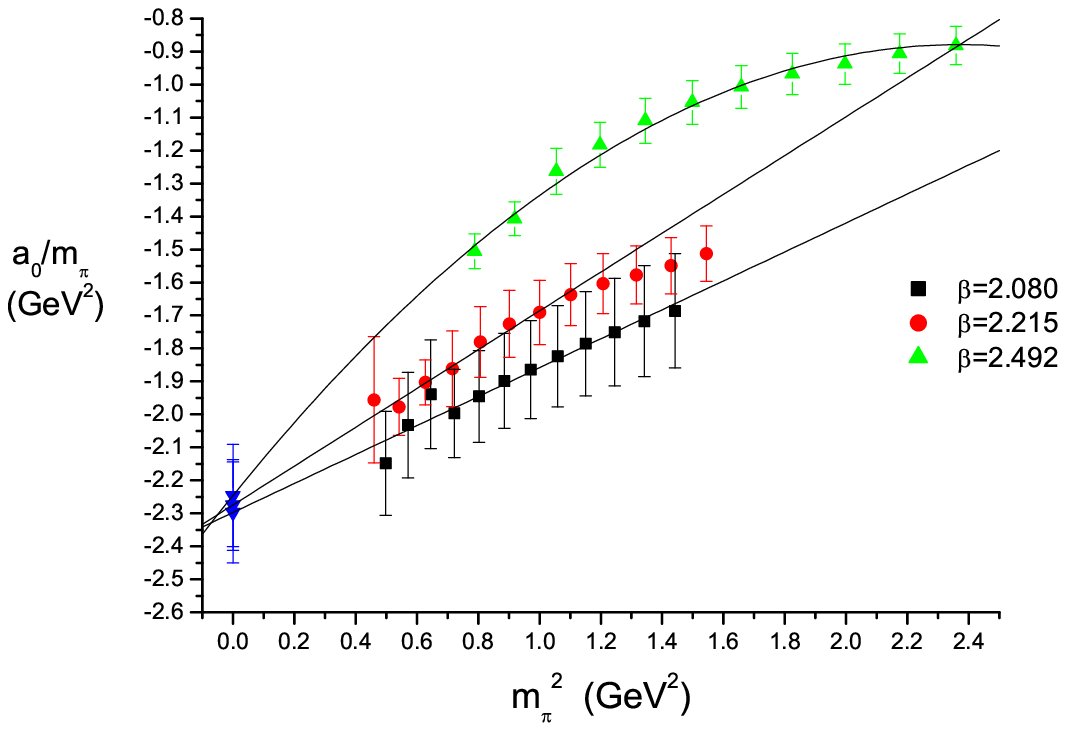}}
 \subfigure{\includegraphics[width=0.48\textwidth]{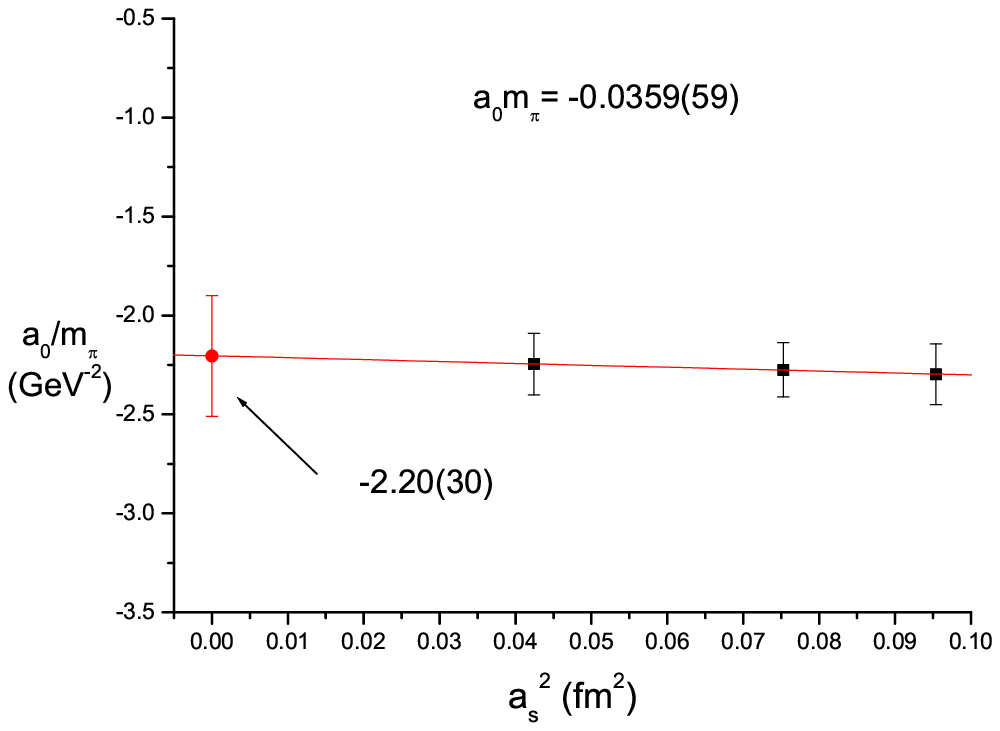}}
 \end{center}
 \caption{Left panel: chiral extrapolation of $\frac{a_0}{m_{\pi}}$
 at $\beta=2.492$ using Scheme 2 while keeping other beta values
 extrapolated in Scheme 1. Right panel:  the corresponding continuum
 limit extrapolation of the quantity $\frac{a_0}{m_{\pi}}$. The
 extrapolated result is indicated by a solid circle near
 $a_s=0$.\label{fig:delta_a0_two}}
 \end{figure}

 \begin{table}
 \caption{Chiral extrapolation fits of $\frac{a_0}{m_{\pi}}$ using
 Scheme 1 and the continuum limit of the fitted parameter $A_0$ (the
 corresponding value in the chiral limit).\label{tab:A0A1_1}}
 \begin{tabular}{|c|c|c|c|c|c|c|}
 \hline $\beta$ & $A_0$ (GeV$^{-2}$) & $A_1$ (GeV$^{-4}$) &
 $\chi^2/d.o.f$
 & range\\
 \hline 2.080 & -2.30(15) & 0.44(16) & 0.05 & 1--12\\
 \hline 2.215 & -2.27(14) & 0.59(17) & 0.04 & 1--8\\
 \hline 2.492 & -2.15(19) & 0.82(20) & 0.09 & 1--4\\
 \hline continuum limit & -2.05(36) & & &\\
 \hline $\chi^2/d.o.f$ & 0.18 & & & \\
 \hline
 \end{tabular}
 \end{table}
 \begin{table}
 \caption{Chiral extrapolation  of $\frac{a_0}{m_{\pi}}$ at
 $\beta=2.492$ using Scheme 2. With others still using Scheme 1, the
 continuum limit of parameter $A_0$ is shown.\label{tab:A0A1_2} }
 \begin{tabular}{|c|c|c|c|c|c|c|}
 \hline $\beta$ & $A_0$ (GeV$^{-2}$) & $A_1$ (GeV$^{-4}$) & $A_2$
 (GeV$^{-6}$)& $\chi^2/d.o.f$
 & range\\
 \hline 2.492 & -2.24(16) & 1.15(21) & -2.24(6) & 0.13 & 1--12\\
 \hline continuum limit & -2.20(30) & & & &\\
 \hline  $\chi^2/d.o.f$ & 0.02 & & & &\\
 \hline
 \end{tabular}
 \end{table}
 We use the quantity
 $a_0/m_{\pi}$, which is finite in the chiral limit,
 for the chiral extrapolation, as suggested by the CP-PACS collaboration~\cite{CPPACS03:pipi_phase}.
 The scale is set using the pure gauge sector with the Sommer scale $r_0=0.5$fm.
 In Chiral perturbation Theory (ChPT),
 The $m^2_\pi$ dependence of the scattering length is known within
 Chiral perturbation theory
 (ChPT)~\cite{gasser-leutwyler:chiral_oneloop_a,gasser-leutwyler:chiral_oneloop_b,colangelo01:pipi}.
 However, it is well-known that
 ChPT is only effective when $m_{\pi}$ is small. The pion mass range
 in our simulation (from 0.7 GeV to 1.5 GeV) is definitely beyond the
 applicability range of ChPT. Therefore, we have attempted to parameterize our
 data for $a_0/m_\pi$ using either a linear function in $m^2_\pi$:
 \begin{equation}
 \frac{a_0}{m_{\pi}}=A_0+A_1m^2_{\pi};,
 \end{equation}
 or a quadratic function in $m^2_\pi$:
 \begin{equation}
 \frac{a_0}{m_{\pi}}=A_0+A_1m_{\pi}^2+A_2m_{\pi}^4\;.
 \end{equation}
 These two methods will be referred to as Scheme 1 and 2.
 It is found that only the data at $\beta=2.492$ show significant
 curvature in the pion mass regime that we are studying.
 For the other two $\beta$ values, quadratic fits do not
 give statistically more favorable results. Therefore, we only
 attempted Scheme 2 for $\beta=2.492$.
 The fitting results in both schemes are tabulated in Table~\ref{tab:A0A1_1} and
 Table~\ref{tab:A0A1_2}. The fittings are also illustrated in
 the left panels of Fig.~\ref{fig:delta_a0_linear} and
 Fig.~\ref{fig:delta_a0_two}.

 Finally, a continuum limit extrapolation is performed to
 get rid of the lattice spacing errors, using a functional form that
 is linear in $a^2_s$. A possible linear term contamination might be
 there but the coefficient of it is too small to be visible in the fitting.
 In Fig.~\ref{fig:delta_a0_linear} and
 Fig.~\ref{fig:delta_a0_two} (right panels), we show the results for the
 continuum limit extrapolation. The straight lines represent the
 extrapolation towards the $a_s=0$ limit and the final results are
 shown as circles. After the chiral and continuum extrapolations,
 our results for the scattering length in this particular channel read:
 \begin{equation}
 a_0m_{\pi}=\left\{\begin{array}{ll}
 0.0399(70)\;, & \mbox{   Scheme 1}\\
 0.0359(59)\;, & \mbox{   Scheme 2}\end{array}\right.
 \end{equation}
 The two results are consistent with each other within
 errors. These results can also be compared with analogous results obtained
 in other theoretical
 calculations~\cite{weinberg66:pipi_CA,bijnens98:chiral_pipi,colangelo01:pipi,zheng05:pipi_dispersion}
 and the experiment~\cite{E86500:pipi}.
 Our results of two schemes are both compatible with the experiment and
 the result of Scheme 1 is
 consistent with the results from other theoretical investigations.
 Our result for the scattering length also agrees with previous lattice results obtained
 by other groups. Table~\ref{tab:a0_summary} summarizes all relevant results for
 the scattering length in this channel.
 \begin{table}
 \caption{A summary of results for the pion-pion scattering
 length.\label{tab:a0_summary}}
 \begin{tabular}{|l|l||l|l|}
 \hline & $a_0m_{\pi}$ & & $a_0m_{\pi}$ \\
 \hline JLQCD (LIN)\cite{jlqcd99:scat} & -0.0406(47) & CP-PACS
 (quenched)~\cite{CPPACS03:pipi_phase} &
 -0.0558(56)\\
 \hline JLQCD (EXP)\cite{jlqcd99:scat} & -0.0410(69) & E865 Collaboration~\cite{E86500:pipi} & -0.036(9)\\
 \hline C.Liu (Scheme 1)\cite{chuan02:pipiI2} & -0.0342(75) & Current algebra~\cite{weinberg66:pipi_CA} & -0.046\\
 \hline C.Liu (Scheme 2)\cite{chuan02:pipiI2} & -0.0459(91) & CHPT (one-loop)~\cite{bijnens98:chiral_pipi}& -0.0423(10)\\
 \hline X.Du\cite{chuan04:pipi} & -0.0467(45) & CHPT (two-loop, Roy Eq.)~\cite{colangelo01:pipi} & -0.0444(10) \\
 \hline CP-PACS (unquenched)~\cite{CPPACS04:pipi_phase_unquench} &
 -0.0266(16)
 & Dispersion relations~\cite{zheng05:pipi_dispersion} & -0.0440(11)\\
 \hline
 \end{tabular}
 \end{table}

 \subsection{The scattering phase shift}

 We now come to the results for scattering phases. Pion
 scattering phases can also be computed in the low-energy regime within
 chiral perturbation theory~\cite{gasser-leutwyler:chiral_oneloop_a,gasser-leutwyler:chiral_oneloop_b}.
 However, as already mentioned above, the formulae thus obtained
 are only applicable for very light pion mass and low scattering
 momenta. We therefore used a
 method that has been used in quenched studies by the CP-PACS
 Collaboration~\cite{CPPACS03:pipi_phase,CPPACS03:pipi_phase_unquench,CPPACS04:pipi_phase_unquench},
 namely we simply parameterize the scattering phases
 with a polynomial in $m^2_\pi$ and the momentum. Our method is
 a modified version of their methods.

 CP-PACS Collaboration~\cite{CPPACS03:pipi_phase,CPPACS03:pipi_phase_unquench,CPPACS04:pipi_phase_unquench}
 defines a scattering amplitude as follows:
 \begin{equation}
 A(m_{\pi},\bar{k})=\frac{\tan\delta(\bar{k})}{\bar{k}}\cdot\frac{E_{\pi\pi}}{2}
 \end{equation}
 where $\bar{k}=|\bar{\textbf{k}}|$. Then, they used a polynomial
 function in both $m_{\pi}^2$ and $\bar{k}^2$ to fit their simulation data.
 However, if some of the phase shifts data exceede the limit $-90^\circ$, since the function
 $A(m_\pi,\bar{k})$ defined above involves the $\tan$ function which
 is discontinuous at $-90^\circ$, this makes the amplitude $A(m_\pi,\bar{k})$
 discontinuous as well which is not  convenient for chiral extrapolations.
 To overcome this difficulty, we
 parameterize the phase shift $\delta$ itself by a polynomial in
 both $m^2_\pi$ and $\bar{k}^2$ as:
 \begin{equation}
 \delta(m_{\pi}^2,\bar{k}^2)=D_{00}+D_{10}m_{\pi}^2
 +D_{20}m_{\pi}^4+D_{01}\bar{k}^2+D_{11}m_{\pi}^2\bar{k}^2+D_{02}\bar{k}^4
 \end{equation}
 \begin{figure}[htb]
 \subfigure{\includegraphics[width=0.48\textwidth]{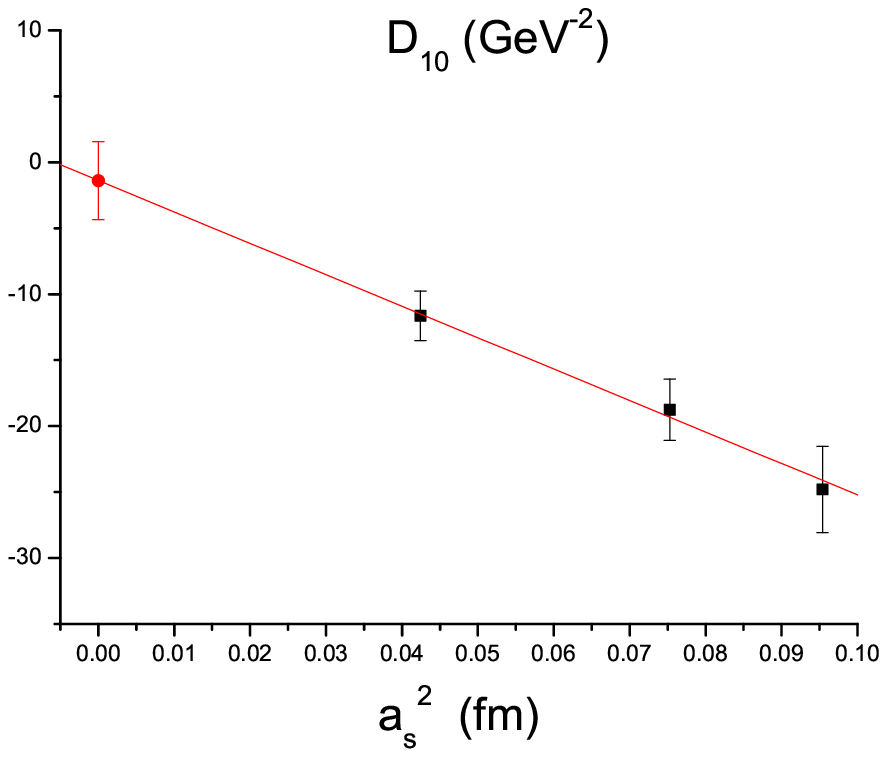}}
 \subfigure{\includegraphics[width=0.48\textwidth]{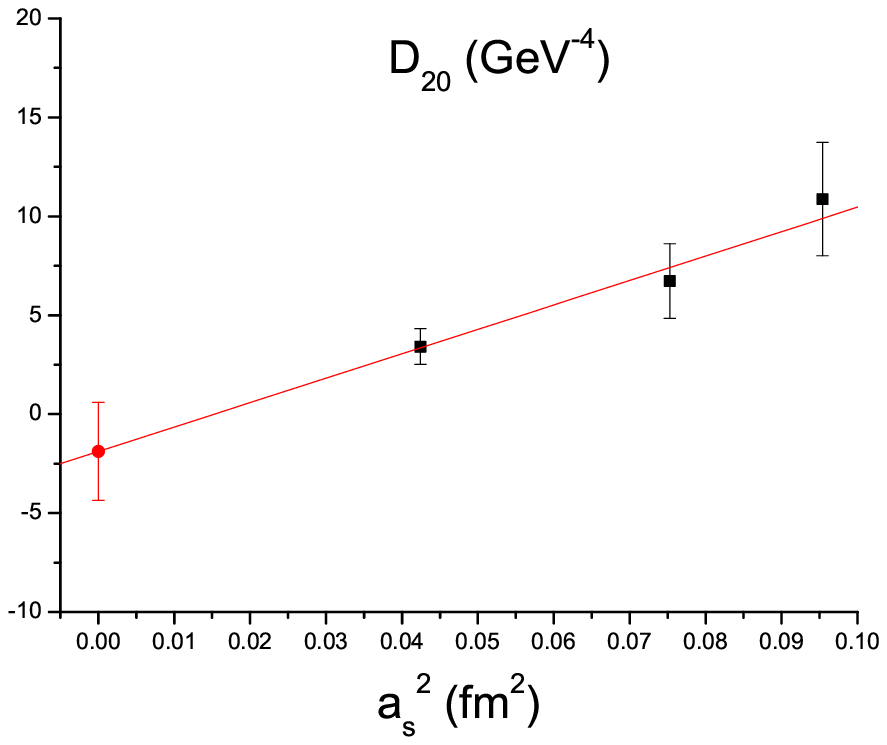}}
 \subfigure{\includegraphics[width=0.48\textwidth]{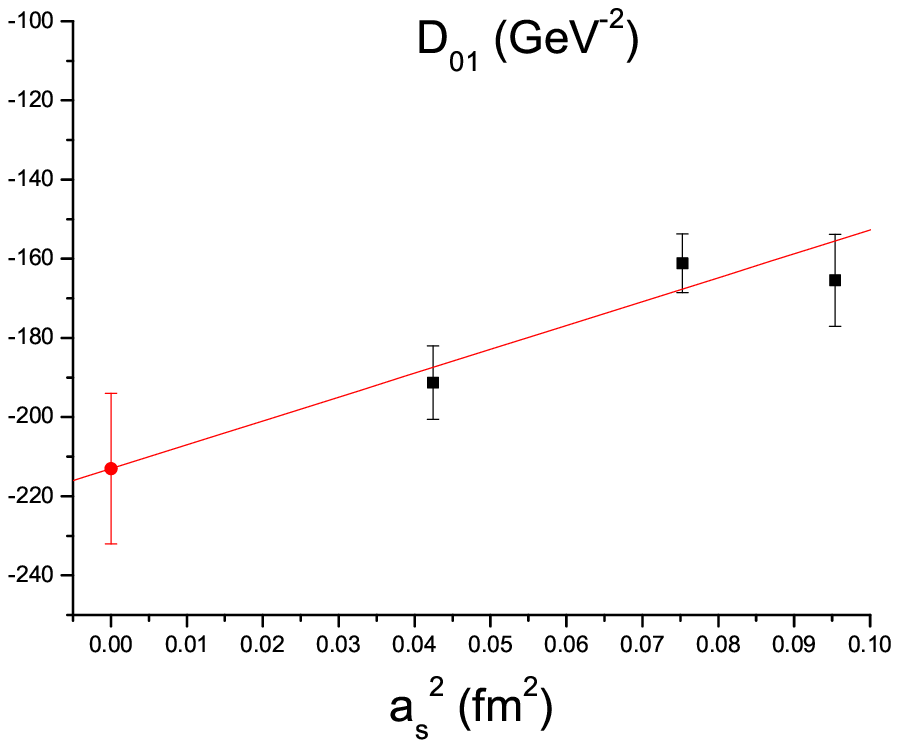}}
 \subfigure{\includegraphics[width=0.48\textwidth]{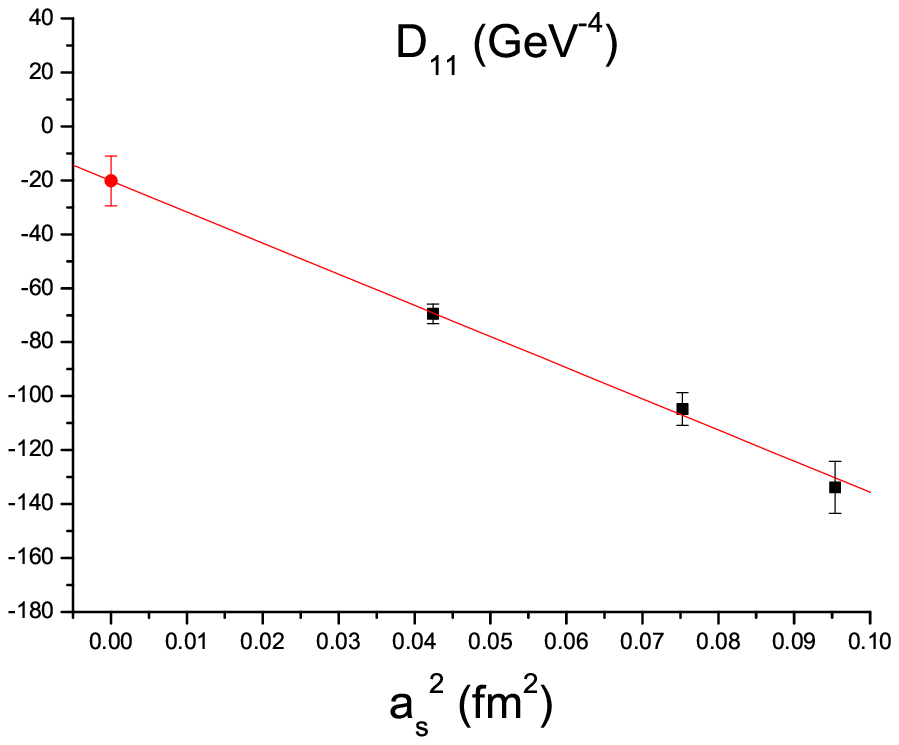}}
 \subfigure{\includegraphics[width=0.48\textwidth]{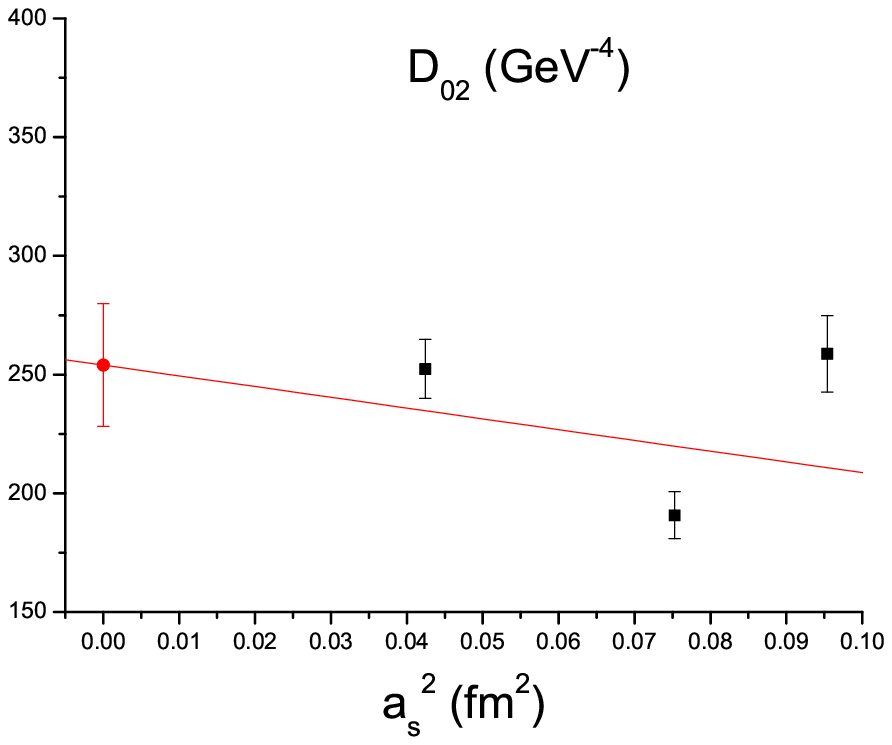}}
 \caption{The continuum limit extrapolation of the parameter
 $D_{ij}$. The solid circles near $a_s=0$ represent the corresponding
 continuum limit values.\label{fig:continuum_Dij}}
 \end{figure}
 The above function includes all terms with powers not larger than four.
 We have tried polynomial functions with higher powers but
 found that they had not improved the fitting quality.
 Note that in quenched lattice QCD, since the chiral behavior is different
 from true QCD, physical quantities can suffer from bad chiral behaviors~\cite{bernard96:quenched_scat}.
 For example, in the above fitting formulae, a non-vanishing constant term $D_{00}$ can
 exist, which would be absent in true QCD due to chiral symmetry.
 In practice, by fitting of our quenched data, we find that $D_{00}$ is
 always consistent with zero within statistical error when we regard it as a free parameter.
 Therefore, in the following discussion $D_{00}$ is fixed to zero.
\begin{table}
\caption{Fitted results for the scattering phase shifts $\delta$ at
each $\beta$. The continuum limit for each parameter is also
shown.\label{tab:delta_beta}}
\begin{tabular}{|c||c|c|c|c|c|c|}
\hline $\beta$ & $D_{10}$ (GeV$^2$) & $D_{20}$ (GeV$^4$) & $D_{01}$
(GeV$^2$) & $D_{11}$ (GeV$^4$) & $D_{02}$
(GeV$^4$) & $\chi^2$/d.o.f\\
\hline 2.080 & -24.8(33) & 10.9(19) & -165(11) & -133.8(96) &
259(16) & 0.98\\
\hline 2.215 & -18.8(23) & 6.7(19) & -161.2(74) & -104.8(61) &
190.8(99) & 2.05\\
\hline 2.492 & -11.6(19) & 3.42(91) & -191.3(92) & -69.5(36) &
252(12) & 0.77\\
\hline  continuum limit & -1.4(41) & -1.9(25) & -213(19) & -20.2(92)
& 254(25) & \\
\hline
\end{tabular}
\end{table}

\begin{figure}[htb]
 \begin{center}
 \subfigure{\includegraphics[width=0.7\textwidth]{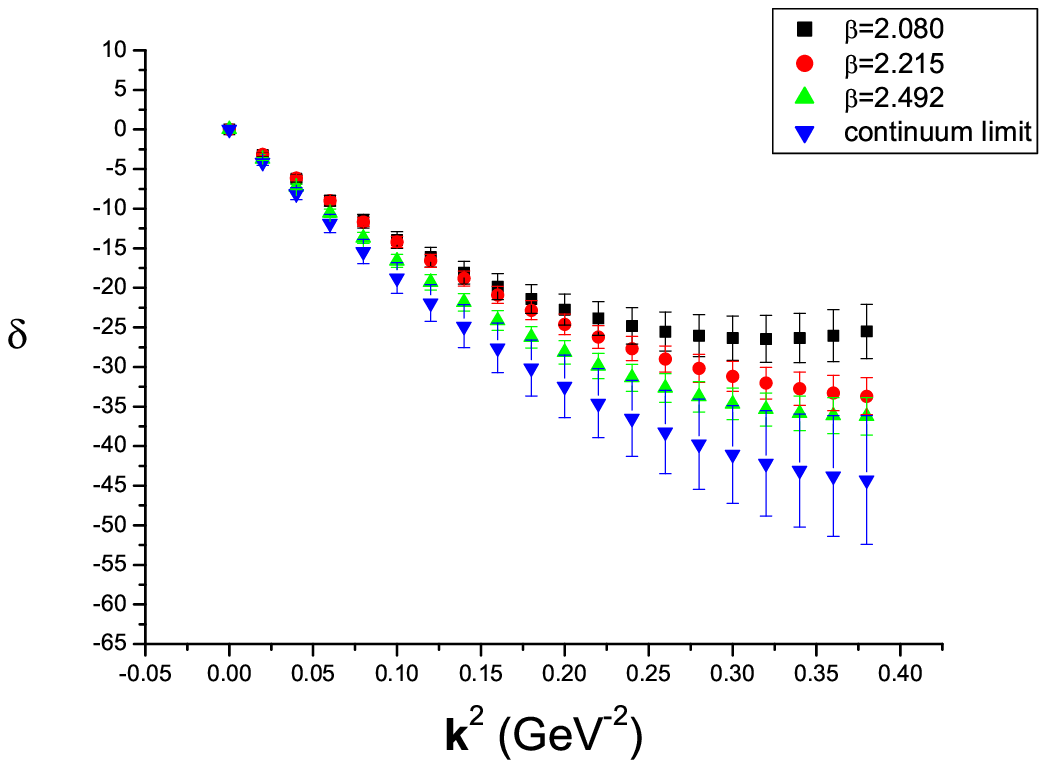}}
 \end{center}
 \caption{The values of the phase shifts $\delta$ after the chiral
 extrapolations at three values of $\beta$. The data points are
 labeled using squares, circles and triangles for three values of
 $\beta$ respectively. Also shown are the continuum limit results of
 the phase shift which are labeled by upside-down
 triangles.\label{fig:continuum_delta}}
 \end{figure}

 All fitting results for $D_{ij}$ are
 tabulated in Table~\ref{tab:delta_beta}. The $\chi^2/d.o.f$ of
 $\beta=2.215$ is somewhat large but still acceptable.
 In Fig.~\ref{fig:continuum_delta}, we plot the fitted results for
 the phase shift $\delta$  as a
 function of momentum $\bar{\textbf{k}}^2$ while setting $m_{\pi}$
 to zero (chiral limit). Results for three $\beta$ values are all shown
 in this figure with different symbols. The results for different lattice spacings
 tend to agree with one another in the low-momentum limit and deviate in
 the large momentum limit as expected. We then perform the continuum limit extrapolation of
 various coefficients $D_{ij}$
 by using a function linear in $a^2_s$. All extrapolations
 tabulated in Table~\ref{tab:delta_beta} are good except for
 $D_{02}$. The continuum extrapolations are also shown in Fig.~\ref{fig:continuum_Dij}.
 After the continuum extrapolation, the results for the phase shift
 as a function of $\bar{\textbf{k}}^2$ are plotted in
 Fig.~\ref{fig:continuum_delta} with upside-down triangles.

 Our results can be compared with previous quenched lattice results from
 CP-PACS collaborations~\cite{CPPACS03:pipi_phase} and unquenched
 results from NPLQCD~\cite{savage06:pipi}.
 We find that they agree with each other within errors.
 Due to the asymmetric box used in this study,
 we are able to compute the phase shifts
 at more values of scattering momentum compared with similar
 calculations using a symmetric box.
 For example, in Ref.~\cite{CPPACS03:pipi_phase}, scattering phases
 are obtained at five values of $\bar{k}^2$ in the range from
 $0.02$GeV$^2$ to $0.34$GeV$^2$. By using an asymmetric box,
 even in a smaller range of $0.02$GeV$^2$ to $0.12$GeV$^2$, we have over
 a dozen of data points for the phase shift which can be
 compared with results from other theoretical investigations
 and the experiments in more detail.

 \begin{figure}[htb]
 \begin{center}
 \subfigure{\includegraphics[width=0.7\textwidth]{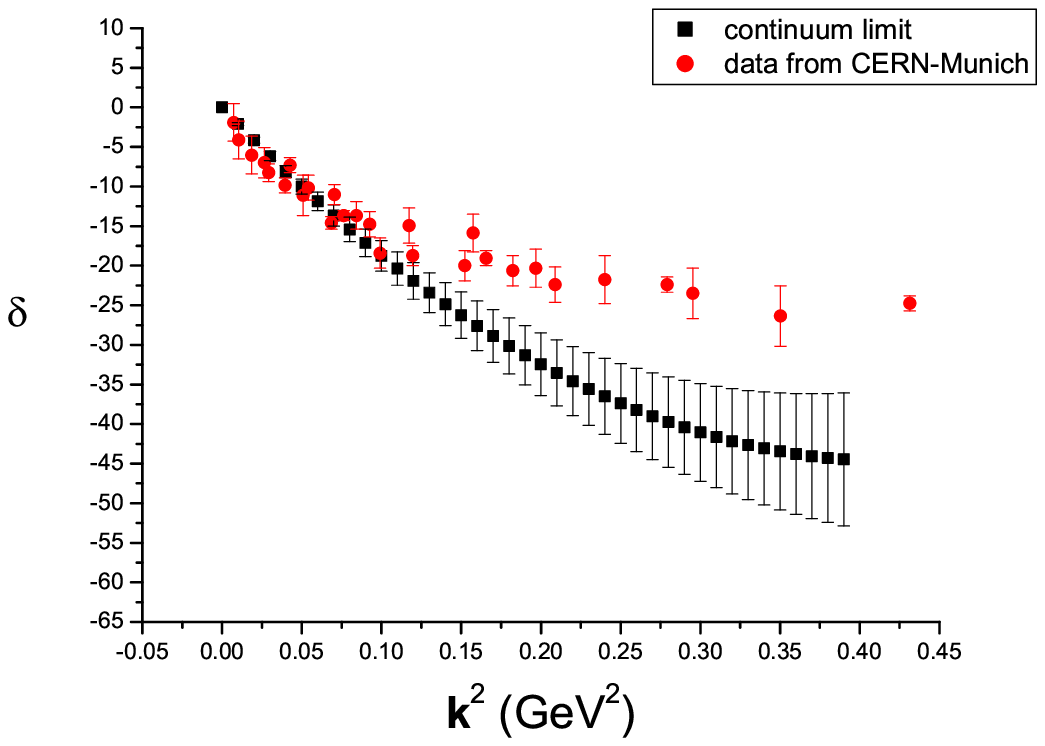}}
 \end{center}
 \caption{Comparison of our lattice results for the scattering phase
 shifts with the experimental data from
 CERN-Munich~\cite{Hoogland:1974cv}. Results are consistent with
 each other for $\textbf{k}^2$ below 0.1 GeV$^2$ which roughly
 corresponds to the center of mass energy of about
 $0.6$GeV.\label{fig:compare_experiment}}
 \end{figure}
 Finally in Fig.~\ref{fig:compare_experiment}, we have shown the
 the same result for the phase shifts $\delta$
 in the continuum limit together with the
 experimental results from CERN-Munich group~\cite{Hoogland:1974cv}.
 It is seen that our final results
 agree with the experimental results within errors for $\bar{\textbf{k}}^2$
 below $0.1$ GeV$^2$ which is about $\sqrt{s}=0.6$ GeV.
 At higher energies, our results deviate from the experimental
 results. This deviation might be caused by the systematic
 uncertainties in our calculation, e.g. quenching and chiral
 extrapolations. Numerically speaking, it is largely due to
 poor determination of the coefficient $D_{02}$.
 However, we would like to point out that, it is clear from our
 quenched calculation that,
 the asymmetric volume technique advocated here would also be
 useful for unquenched studies once
 the unquenched configurations become available.

\section{Conclusions}

 In this paper, we propose to study hadron-hadron scattering processes on
 lattices with asymmetric volume. This setup has the advantage that it provides
 much more non-degenerate low-momentum modes with a relatively small volume, allowing
 more detailed comparison both with the experiments and with other theoretical results.
 To illustrate the feasibility of this proposal,
 pion pion scattering length and scattering phases
 in the $I=2$, $J=0$ channel are computed within quenched
 lattice QCD using clover improved lattice actions on anisotropic lattices.
 Our quenched results indicate that the usage of asymmetric volumes
 indeed allow us to access much more low-momentum modes than in the
 case of cubic volume of similar size.
 For $\bar{k}^2$ in the range  of $0.02$GeV$^2$ to $0.12$GeV$^2$, we have over
 a dozen of data points for the phase shift, much more than that in
 the cubic case with similar volume. It is also noted that, in the low-momentum
 region, after the chiral and continuum extrapolations, our results for the
 scattering length and the scattering phase shifts are in
 good agreement with the experimental data and are consistent with
 results obtained using other theoretical means.

 Although our calculation is now performed in the quenched approximation,
 similar calculations are also possible in the unquenched case once the gauge
 field configurations become available.
 Finally, we have only computed scattering length and phases
 in the $I=2$, $J=0$ channel. Phenomenologically speaking, other
 channels, in particular $I=J=0$ channel, are more interesting.
 However, this channel is difficult for two main reasons: One
 needs to perform a full QCD calculation otherwise the theory
 is sick in the chiral limit~\cite{bernard96:quenched_scat};
 one has to deal with vacuum (disconnected) diagrams
 which significantly increase the amount of computational cost.
 Also interesting and equally challenging is the $I=J=1$ channel
 where one would expect to see a rho resonance~\cite{CPPACS06:pipi_rho}.
 We expect the use of asymmetric volumes should also be
 beneficial in these studies since a lot more low-momentum modes
 become accessible in an asymmetric box.

 \section*{Acknowledgments}
 The authors would like to thank Prof.~H.~Q.~Zheng and
 Prof.~S.~L.~Zhu of Peking University for helpful
 discussions. The numerical calculations were performed on DeepComp 6800 supercomputer of the
 Supercomputing Center of Chinese Academy of Sciences, Dawning 4000A
 supercomputer of Shanghai Supercomputing Center, and NKstar2
 Supercomputer of Nankai University.

 \bibliography{PiPi_phase_shortened}

\end{document}